\documentclass[conference]{IEEEtran}
\IEEEoverridecommandlockouts
\usepackage{cite}
\usepackage{amsmath,amssymb,amsfonts}
\usepackage{algorithmic}
\usepackage{graphicx}
\usepackage{textcomp}
\usepackage{subcaption}
\usepackage{svg}
\usepackage{graphicx}
\DeclareGraphicsExtensions{.pdf,.jpeg,.png,.svg}
\usepackage{caption}
\usepackage{textcomp}
\usepackage[hyphens]{url}
\usepackage[hidelinks]{hyperref}
\hypersetup{breaklinks=true}

\usepackage{pgf}
\usepackage{siunitx}

\usepackage{xcolor}
\def\BibTeX{{\rm B\kern-.05em{\sc i\kern-.025em b}\kern-.08em
    T\kern-.1667em\lower.7ex\hbox{E}\kern-.125emX}}

\begin{document}
%
\title{A Layered Architecture Enabling Metaverse Applications in Smart Manufacturing Environments 
}

\author{
    \IEEEauthorblockN{Armir Bujari, Alessandro Calvio, Andrea Garbugli, Paolo Bellavista}
    \IEEEauthorblockA{
        University of Bologna\\
        Department of Computer Science and Engineering\\
        Bologna, Italy \\
        name.surname@unibo.it}
}

\maketitle

\begin{abstract}
The steady rollout of Industrial IoT (IIoT) technology in the manufacturing domain embodies the potential to implement smarter and more resilient production processes. To this end, it is expected that there will be a strong reliance of manufacturing processes on cloud/edge services so as to act intelligently and flexibly. While automation is necessary to handle the environment's complexity, human-in-the-loop design approaches are paramount. In this context, Digital Twins play a crucial role by allowing human operators to inspect and monitor the environment to ensure stability and reliability. Integrating the IIoT with the Metaverse enhances the system's capabilities even further, offering new opportunities for efficiency and collaboration while enabling integrated management of assets and processes. This article presents a layered conceptual architecture as an enabler for smart manufacturing metaverse environments, targeting real-time data collection and representations from shopfloor assets and processes. At the bottom layer, our proposal relies on middleware technology, serving differentiated Quality of Service (QoS) needs of the Operation Technology (OT) monitoring processes. The latter contributes to feeding a virtual layer where data processes reside, creating representations of the monitored phenomena at different timescales. Metaverse applications can consume data by tapping into the metaverse engine, a microservice-oriented and accelerated Platform as a Service (PaaS) layer tasked with bringing data to life. Without loss of generality, we profile different facets of our proposal by relying on two different proof-of-concept inspection applications aimed at real-time monitoring of the network fabric activity and a visual asset monitoring one.   

\end{abstract}


\section{Introduction}

We are in the middle of an industrial revolution, i.e., of Industry 4.0 (I4.0)-based intelligent and cooperative Cyber-Physical Systems (CPSs), underpinned by a digital transformation that will affect all industries. At the core of this revolution is the convergence of flexible and multi-faceted communication technologies, novel computing techniques such as cloud/edge computing, and the application of the IoT vision to industrial manufacturing systems~\cite{zvei}. As a result, industrial devices and machines will rely on heterogeneous wireless/wired technologies to communicate with applications running on global cloud/local edge platforms, thus enabling new efficient solutions, e.g., for predictive maintenance and process optimization.

The application of big data and AI has been successfully adopted to streamline and optimize non-manufacturing processes and is now being expanded into the industrial sector. Today's manufacturing environment involves machines that can communicate independently and generate data at a rapid pace. This information can be utilized proactively to enhance control, and business processes in manufacturing, engineering, supply chain, and product life cycle management~\cite{wef}. A significant challenge in realizing this goal is the outdated and inflexible division between technology departments involved in product manufacturing and those focused on management tasks. Indeed, industrial automation has taken a traditional approach, opting for a strict separation between Operation and Information Technology domains (OT \& IT)~\cite{gartnerOTIT}. 

Until now, IT technologies such as cloud/edge computing, Service-Oriented Architectures (SOA), and virtualization have been used in the industrial domain in limited ways, only where stringent requirements were not necessary \cite{pop}. However, it is becoming clear that smart manufacturing environments will have a significant impact only with a complete convergence of OT/IT, allowing for full utilization of data intelligence and recent computing and communication technologies.

While recognizing the OT/IT synergy as a major milestone in tackling the complexity of this evolving environment, human-in-the-loop design approaches are paramount. Despite the many benefits that data intelligence can provide to society and the economy, the technology behind it can also have negative or unpredictable impacts, bringing about new risks to individuals. A synergetic development of both AI-driven and human-centric approaches is key to driving a safe, resource-efficient, and sustainable manufacturing domain~\cite{EUI5.0}. 

To this end, a central component toward this objective is the concept of the digital twin; a virtual replica of a physical asset that can be monitored and optimized in real-time. Data streams often connect this digital replica and the physical counterpart, feeding and continuously updating the digital model, used as a descriptive or predictive tool for planning and operational purposes \cite{grieves}. Integrating the shopfloor with the Metaverse enhances the system’s capabilities even further, offering new opportunities for efficiency and collaboration while enabling integrated management of assets and processes \cite{metaverseDef}. In a Metaverse, digital replicas of real-life objects and processes can be used for various purposes such as testing, monitoring, and collaboration through AR/VR technology. 

In this work, we present a layered conceptual architecture and accompanying software prototype conceived to enable a human-in-the-loop manufacturing environment enriched thanks to real-time data sourced from shopfloor processes. Data are stored at different layers of the architecture, having different scopes, and are brought to life via virtual representations of assets, depending on the layer and the application used to access it. In synthesis, the architecture consists of multiple layers, including a bottom layer that utilizes middleware technology to serve the different Quality of Service (QoS) requirements of the OT monitoring processes, implementing a beyond state-of-the-art OT/IT convergence mechanism. 

Moving up is the virtual layer tasked with the creation of multi-faceted representations of the monitored phenomena at different timescales. Herein are logically collocated the repositories containing the virtual representations of assets at different levels of granularity, e.g., individual machines or production cells within the shopfloor, etc. Finally, at the application layer reside the immersive and collaborative applications. The applications leverage the metaverse engine, a microservice-oriented and accelerated Platform as a Service (PaaS) layer that fuses and renders synchronized information sources from the virtual layer. Without loss of generality, we profile different facets of our proposal by relying on two different proof-of-concept inspection applications aimed at real-time monitoring of the network fabric activity and a physical asset monitoring one.


\section{Background}
\label{background}
In this section, we provide a concise survey on the paradigm shift advocated by Industry 4.0, motivating the need for a human-centric approach, as recently advocated by many nation-state initiatives on their next industrial vision. Next, we provide a concise survey on the technology behind some of the adopted design choices.

\subsection{Towards a Human-in-the-loop Manufacturing Environment}

\begin{figure}[t]
  \centering
  \includegraphics[width=1\columnwidth]{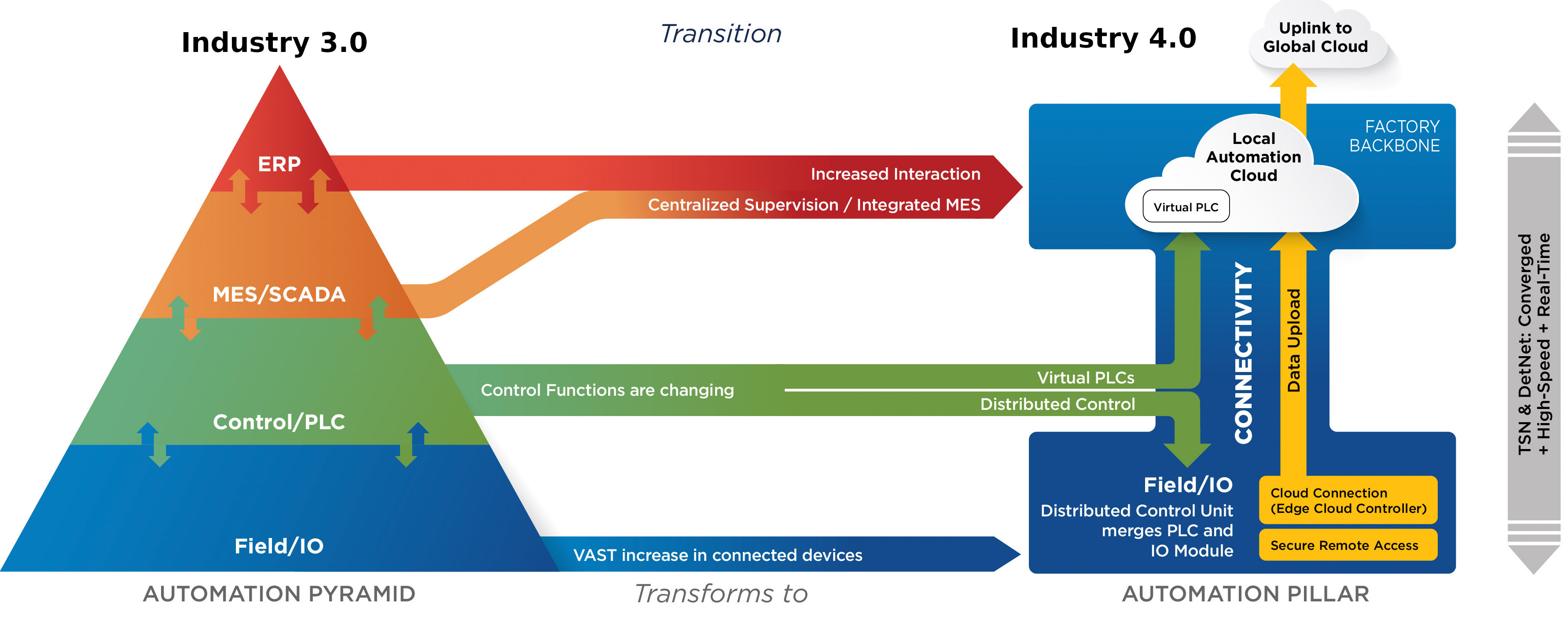}
  \caption{Transition to Industry 4.0.}
  \label{fig:transition3to4}
\end{figure}

The automation pyramid serves as a reference model for dealing with the challenges posed by complex and heterogeneous manufacturing systems and consists of different levels, each representing a specific stage of automation \cite{pyramidModel}. 
A simplified specification consists of four different hierarchy levels (see left part of Fig.~\ref{fig:transition3to4}). 

The sensing and actuation units are located on Level 1. They commonly operate within milliseconds or seconds to control the physical manufacturing processes to reach the required quality demands. Monitoring, supervision, and control is the task of Level 2 systems. Production environments mainly use programmable logic controllers (PLCs). Depending on the individual production system, PLCs operate within hours down to less than periods of seconds. As Level 2 directly controls the actuators, PLCs embedded logic may implement short-term adaptions of the process(es).

Level 3 of the automation pyramid encompasses activities related to the management and control of manufacturing operations. This includes dynamic scheduling of jobs, optimization of production processes, and data aggregation and distribution, all encompassed in a Manufacturing Execution System (MES), which operates within one day or shift and must be capable of reacting to unexpected events, e.g., machine breakdowns. The level also includes SCADA systems that collect real-time data, data-driven analysis, and decision support systems to optimize processes. This level bridges lower levels of control and monitoring and higher levels of strategic decision-making and planning.

The top Level 4 consists of an Enterprise Resource Planning (ERP) tool, which executes the required tasks and manages inventory and resources. As the ERP defines the long-term production utilization, the long-term operations (optimizations) are also determined here.
 
Cyber-physical systems, which merge the physical and virtual world through embedded hardware and software systems, present a new, non-hierarchical approach to production (see right part of Fig.~\ref{fig:transition3to4}). This change is due to the need to better highlight and distinguish the networks that connect the layers and to reflect that some technologies may no longer reside in the facility but in local/remote edge-cloud computing environments. This has been made possible thanks to the convergence of the so-called disaggregation trend in the IT industry and the market penetration of deterministic and high-throughput communication technologies such as Time Sensitive Networking (TSN) and 5G~\cite{tsn,tsn5g}. In this \emph{automation pillar}, processes that do not require hard, real-time control can now be run in the cloud with virtual PLCs, as shown in the graphic. The graphic also depicts how, in the IIoT, data can now be shared more easily through all levels rather than sequentially from one layer to the next. 

The hierarchical architecture of the automation pyramid is still present and very common in production systems. This is also indicated by the adoption of the automation pyramid within the more recent RAMI 4.0 reference model for Industry 4.0~\cite{rami}. However, the paradigm shift has already begun by recognizing the OT/IT convergence problem, paving the way for the use of descriptive and predictive Digital Twin technology in manufacturing systems. The availability of data and recent advances in AI technology could pave the way for smarter manufacturing environments, but this evolution should go in pace with a human-centric design approach. In this context, the notion of the Metaverse could enhance an operator's capabilities even further, offering new opportunities for efficient collaboration while enabling integrated management of assets and processes in the manufacturing domain. Our proposal goes in this direction, providing the basic building blocks for enabling an immersive, human-in-the-loop manufacturing environment.

\subsection{Data Homogenization and Transport from the Shopfloor}

The OPC Unified Architecture (OPC UA) is an industrial automation standard that aims to solve the issue of incompatibility among different Ethernet technologies (such as PROFINET, EtherCAT, and Modbus-TCP) by providing a unified and consolidated view of assets and processes~\cite{opc1,opc2}. This platform-independent standard can facilitate interoperability among vendors.

The initial OPC UA standard employed a Client/Server paradigm: an OPC UA server offers access to data and functions structured in an object-oriented information model, where clients interacted with the information model through standardized services. Based on the request-response method, this communication model does not satisfy the needs of some industrial (control) applications as it creates a strong coupling between different system components and does not meet the performance demands of (hard) real-time systems.

To overcome these limitations, Part 14 of the OPC UA specification introduces an extension based on the Publish/Subscribe (Pub/Sub) communication paradigm~\cite{opc3}. In this model, an application can act as either a publisher or a subscriber (or both), where the publisher sources the data, and the subscriber consumes it. Communication between publishers and subscribers is message-based, with publishers sending messages to a message-oriented middleware without prior knowledge of the subscribers. The latter express interest in certain types of data without having specific knowledge of the publishers. This message-based communication is ideal for applications that value location independence and scalability.


OPC UA is a significant IIoT protocol for addressing communication needs at the OT layer, but it does not fully meet the requirements of the IT layer. Here, there is a need for solutions and frameworks that can handle high-volume data transfers in a secure and reliable manner, which are not priorities of the OPC UA standard. To meet these requirements, we rely on Apache Kafka, an open-source Message Oriented Middleware that uses a Pub/Sub communication model enabling a many-to-many communication pattern~\cite{kafka}. Kafka is well-suited for scenarios where scalable and loosely-coupled systems must work together. In this setting, producers publish messages or batches of messages on a channel called topic, while consumers can read the messages by subscribing to a specific topic. Topics in Kafka make use of partitions, which can be thought of as an infinite log file that is not immediately flushed to disk, leading to highly efficient I/O messaging and having the capability to force strong ordering constraints on message delivery. 

In the following, we present a detailed overview of the proposed layered architecture, discussing some design choices and technological building blocks.
\section{A Layered Architectural Approach}
\label{proposal}
\begin{figure}[t]
  \centering
  \includegraphics[width=0.95\columnwidth]{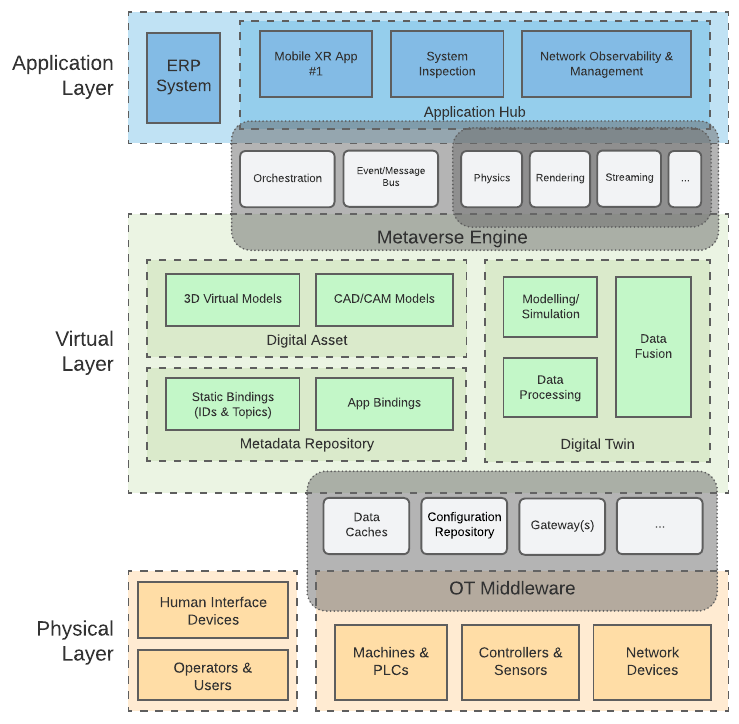}
  \caption{Proposed conceptual architecture.}
  \label{fig:our-architecture}
\end{figure}

The conceptual architecture, depicted in Fig.~\ref{fig:our-architecture}, takes a structured approach to integrate the virtual and the physical realms of next-generation manufacturing systems. In the following, we provide an in-depth description of the various layers, discussing their functional components and some concrete technological choices.

\subsection{Operational Technology Layer}

The \emph{Physical layer} encompasses Level 1 of the automation pyramid, including controllers, machinery, sensors, and processes that make up the industrial shop floor. In addition to these physical components, the physical layer includes users and human operators. Moreover, the layer includes various interfaces, such as haptic technology, VR/AR headsets, and touchscreens, enabling interaction between the physical and the virtual world.

One of the key aspects of our proposal is the implementation of a scalable and reliable solution addressing OT/IT convergence. This convergence is made possible by relying on an OT middleware, which serves as a bridge to gather and elaborate data before they are sent to the virtual layer. One of the main problems when dealing with the convergence of OT and IT is that OT systems often employ a variety of communication protocols, as previously discussed. Therefore, the role of the OT middleware is to standardize the data collection process, abstracting from the details of the specific protocol. 

Another common challenge is data format heterogeneity and resource configuration modeling. We address these challenges by using a standard and common information model (object model) available in the OPC UA dictionary. We anticipate our testbed uses a deterministic network fabric (TSN), and communication on the shopfloor relies on the OPC UA PubSub profile.

As part of the OT middleware, one or more Gateway components are responsible for listening to OPC UA PubSub endpoints and managing data flows between the physical and virtual layers. The Gateway uses different topics and partition levels to prioritize specific traffic flows, such as monitoring data or controlling data traffic in the network. To increase the reliability of the data collection process, the Gateway implements a replication mechanism. For example, monitoring and control topics are assigned a single partition with a high degree of replication. In contrast, data topics from raw sensor elements are assigned multiple partitions with a lower degree of replication.

The OT middleware is also responsible for forwarding the data to the IT level, which logically spans across the upper layers of the architecture. To achieve this, our solution relies on Apache Kafka, a message-oriented middleware, well-suited for scenarios where scalable and loosely coupled systems must interoperate.

\subsection{Virtual Layer}
\label{sec:virt-layer}

The virtual layer is conceptually located above the physical layer and serves as the metaverse engine's primary data access layer and support component. It integrates essential functions utilized by the upper layers in the construction of virtual worlds. Going into more detail, the virtual layer is composed of three main sub-blocks: \emph{digital twin}, \emph{digital assets}, and \emph{metadata repository}.

\subsubsection{Digital Twin}
The digital twin subcomponent is a crucial part of the virtual layer in the factory environment. It creates and maintains digital twins of all the physical machines, products, and facility assets. This allows for a comprehensive representation of the factory's operations.

In addition to creating digital twins of individual assets, this subcomponent maintains a global view of the facility, including both the plant and the network. This allows for a holistic understanding of the entire factory's operations and enables the identification of interdependencies and bottlenecks in the production process.

Here, the processing module collects data from the lower layers and transforms it with varying spatial and temporal resolution levels, e.g., allowing for a more detailed and accurate representation of the physical assets and processes.
For example, the processing module can collect and aggregate data at a high pace, providing real-time monitoring of machine performance. On the other hand, it can be dynamically reconfigured to adjust the scope of the data forwarding processes, feeding data from single production units to an entire production cell, and providing a detailed representation of current processes in the factory.

Combining the various data perspectives generated by the processing layer to create a unified and multi-faceted view of the operations within the factory can represent a challenge. To overcome this, the data fusion layer utilizes various techniques, such as statistical, rule-based, and ontology-aided methods, to synthesize the data into a single, consistent representation of a particular phenomenon. For example, data from different parts of a machine or the machine and its products can be combined to identify the root cause of process issues or inefficiencies. Furthermore, simulation engines are used in this layer to perform several kinds of analysis to gain deeper insights into the potential future evolution of the (sub)systems.


\subsubsection{Digital Asset}
This part consists of an indexed repository, which includes digital representations of physical assets. This data encompasses a wide range of information, including geometric structures, physical attributes, and technical specifications, and may consist of both structured and unstructured information. The current prototype employs various storage technologies to enable data injection into the data lake to promote efficient data management. 

For instance, we rely on a NoSQL database technology such as MongoDB to store device configurations or other static information. In contrast, data sourced from the shopfloor are stored in time-series databases such as Prometheus. Additionally, object storage solutions such as MinIO or Ceph are deployed for storing digital resources such as CAD/CAE and 3D models and multimedia data (i.e., audio/video files).

\subsubsection{Metadata Repository}
An important data repository that allows the association of different sources of information, at different levels of spatial and temporal granularity, between objects. Specifically, the \emph{Static Binding} stores the binding between the data sources (Kafka topics) and the actual assets generating them. To uniquely identify assets and their local in the RAMI 4.0 hierarchy model, we adopt IEC 62264 identifiers. This allows us to keep track of assets and their deployment context, e.g., the production cell where they are located. Another essential piece of information in this hierarchical binding is the Kafka topic, a human-readable string associated with the asset identifier. 

Last is the \emph{App Bindings} module, which stores information about the resources that an application can access, thus allowing the retrieval of data streams related to a particular resource and its model from the \emph{Digital Asset} repository and to overlay this information on the application itself.
This wiring presents a flexible structure, allowing Metaverse applications to interact with objects in space, performing zoom-in and out operations while retaining context and overlaying real-time data of interest.        

\subsection{The Metaverse Engine}
\begin{figure}[t]
  \centering
  \includegraphics[width=0.95\columnwidth]{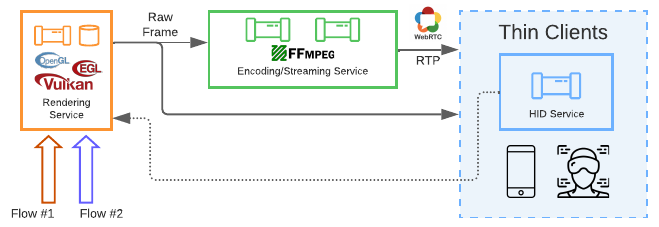}
  \caption{An example of a service chain that utilizes the metaverse engine, where the rendering service acquires data inputs in the form of a time-series data which are superimposed on a video stream (Flow \#1), spatial contextual information (Flow \#2), and user inputs. The service either forwards the raw frame directly to the client so as to visualize the scene through an AR headset or employs a supplementary service for encoding and transmitting the frame using widely adopted streaming protocols (e.g., RTP, WebRTC).}
  \label{fig:engine-subsystems}
\end{figure}

This component acts as a bridge between the virtual world and the application layer. It comprises a series of subsystems typically found in engines for augmented or virtual reality experiences and interactive applications. These subsystems, depicted in Fig.~\ref{fig:engine-subsystems} include physics simulators, rendering engines, streaming services, and other similar components deployed as a series of microservices that can be combined to form more complex services or entire applications. To ensure that each service can meet its QoS/E requirements, the engine can be replicated and distributed across multiple infrastructure nodes.

The second part of the Metaverse Engine is a resource-aware orchestrator that enables the deployment of the microservices and infrastructure components based on the requirements of individual microservices~\cite{refGoodIT}. The orchestrator leverages a deployment strategy that maps service requirements to the underlying resources and guarantees end-to-end QoS/E specifications of various applications. For instance, in the case of shared applications, where multiple users can interact with the virtual world simultaneously, the services can be associated with groups of players to ensure a better overall experience.

The engine has synchronization and state management mechanisms to support various deployment options. The synchronization mechanism manages interactions between the metaverse engine services and the virtual world services, as well as between platform users and these services. The state manager ensures that the consistency of the state is maintained, both for individual applications and for the various microservices, especially in the case of their replication across multiple infrastructure nodes.

The engine is underpinned by a communication bus that enables the exchange of data between the engine's microservices and the applications running within the platform through an event or message system. For instance, a rendering service may be notified of modifications to a scene following a simulation executed by a physics service. The communication bus can utilize different communication and acceleration or memory/storage access technologies, based on node availability and the requirements of individual services, in a manner that is transparent to them. On the network communication side, the bus can utilize technologies such as DPDK, and RDMA, rely on the classic TCP/IP network stack, or IPC mechanisms such as shared memory~\cite{insanePoster}. Concerning memory/storage access, the engine can employ acceleration technologies such as SPDK or GPUDirect to enhance access and transfer of large data and models. These elements and mechanisms, such as orchestration, synchronization, and the communication bus, are crucial to support the application hub model (Sec.~\ref{sec:app-layer}), where different metaverse applications can coexist and be utilized seamlessly by users and operators of the smart factory.

\subsection{Application Layer and Use-cases}
\label{sec:app-layer}

The central component of the application layer is the application hub, a logically centralized repository of applications that can exploit the metaverse engine. The hub can be accessed by industrial plant operators, offering a user-friendly interface that facilitates interaction with the metaverse engine and the ability to switch between applications seamlessly. A notable feature of the hub and its applications is the capability to display information selectively in the user interface, such as runtime data streams and network performance. The application context is established at application design time, and the state is stored and retrieved from the \emph{App Bindings} available at the Virtual Layer.

Furthermore, the hub grants access to individual machines and controllers, enabling real-time monitoring of ongoing production processes. In essence, the hub and its applications serve as an \emph{enriched} access point to the virtual layer and its subcomponents outlined in Sec.~\ref{sec:virt-layer}.

\begin{figure*}[ht]
  \centering
  \includegraphics[width=0.9\textwidth]{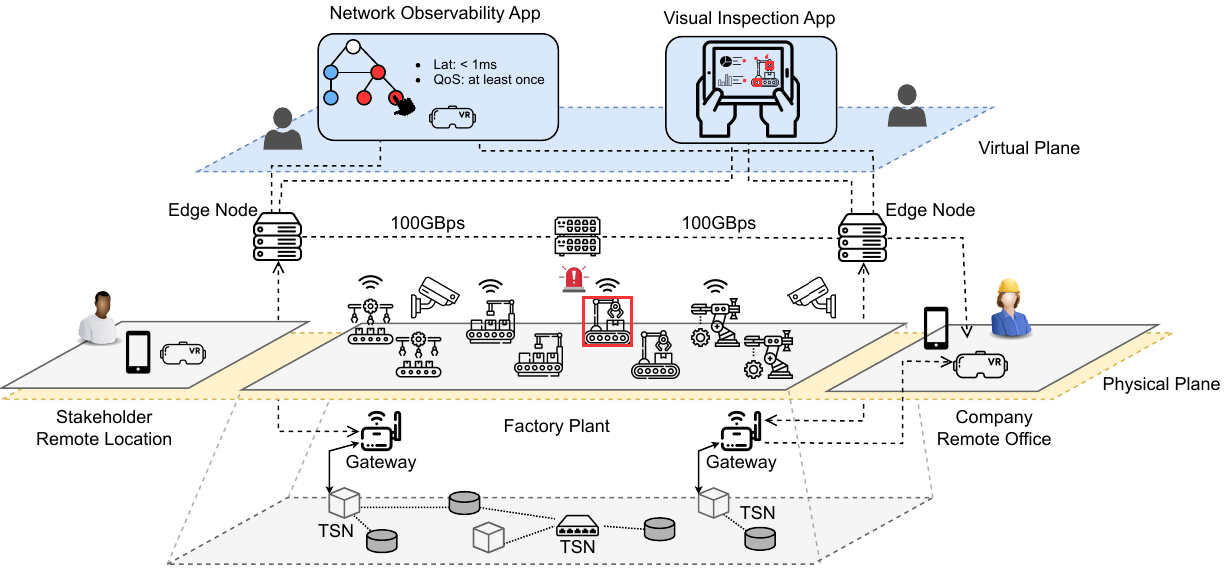}
  \caption{The image represents a smart factory where data sourced from various machines and processes is collected to feed the Virtual Plane. The digital twins are local instances deployed at edge/cloud nodes and can be transferred across them on demand. The upper layer shows a representation of two virtual applications: (i) a network (observability) and (ii) a mixed reality app. used to inspect potential problems on a shopfloor machine.}
  \label{fig:extension_architecture}
\end{figure*}

Fig.~\ref{fig:extension_architecture} shows a simplified instance of the proposed architecture for two use cases to understand better how applications interact with the metaverse engine. The first proof-of-concept application is an AR solution for visualizing the real-time network fabric servicing control and best-effort flows on the shopfloor network. The app. provides the means to monitor the behavior of the network and/or appliances, overlaying and updating the rendered image via real-time data from the control plane. The pictorial representation showcases a graph-like structure, displaying the different data flows, such as control and sensor data. Visualizing the network in an AR environment allows operators and maintenance personnel to quickly identify and understand the relationships between various devices and data flows. This can lead to improved decision-making and increased operational efficiency, reducing downtime and improving the overall performance of the critical industrial network. In addition to visualizing the network fabric and real activities therein, the application provides the means for operators to manage the network in real-time. 

The second proof-of-concept application aims to enhance the efficiency and accuracy of maintenance tasks in industrial settings through video streaming rendered through an augmented reality (AR) appliance. The solution involves streaming high-resolution video data from industrial cameras to an edge node for further processing. The processed video is then transmitted to an AR headset worn by an operator performing maintenance on industrial shopfloor appliances. AR technology enhances the operator's situational awareness by providing real-time, hands-free access to machine information and data. Operators can visualize machine components and systems in real-time, reducing the risk of human error. 
As an example, in Fig.~\ref{fig:extension_architecture} is shown the visual inspection application which is deployed at the edge, receiving and processing the video frames acquired by the industrial camera(s) deployed on the shop floor. The metaverse engine and optimized middleware support ensure that the video is of sufficient quality for display via the AR headset. In this scenario, both video and metric data sourced by the digital twin of the asset are combined so to show a more representative view of the phenomenon. 

\section{Preliminary Evaluation}
\label{evaluation}
This section presents a preliminary evaluation of our proposal in a teal testbed. We first discuss the experimental settings and successively present the obtained results. 

\subsection{Settings}

The deployment environment consists of five nodes, each serving different functionalities within the OT and IT layers. The IT layer includes the virtual and application layers. The OT layer comprises three interconnected nodes, linked by a mesh topology using 1Gbit links. Each OT node is equipped with an Ubuntu 22.04.1 LTS operating system, an Intel Core i5-2400 CPU @ 3.10GHz processor, and 8GB of RAM. Nodes 1 and 2 are dedicated to traffic simulation, utilizing software packages to emulate realistic industrial machine traffic, as described in a previous work~\cite{bosi2019cloud}. On the other hand, Node 3 hosts a Gateway entity, acting as a bridge between the OT and IT layers. The edge nodes, Node 4 and Node 5 share the same operating system and have an 18-core Intel i9-10980XE CPU @ 3.00GHz, along with 64GB of memory. They are directly connected via two 100Gbps Mellanox DX-6 NICs, ensuring minimal network overhead. Additionally, Nodes 4 and 5 serve as the deployment location for the IT subsystems.

The first scenario involves transmitting operational data from a simulated industrial asset at Node 1 via the OPC UA PubSub protocol to the OPC UA Subscriber at Node 2. This scenario simulates a typical sensor-to-controller scenario in an industrial environment, where the internal operational status of the industrial asset is extracted and transmitted using the OPC UA protocol. The Gateway component at Node 3 subscribes to these messages (OPC UA PubSub), then transmits them to the Kafka deployment at Node 4, which the custom Kafka consumer will eventually receive at Node 5. 

The second scenario focuses on simulating the streaming of a dedicated video surveillance system designed to monitor a workshop. Within this scenario, the rendering and streaming services of the Metaverse Engine are deployed on Node 4, while a client application for receiving the video stream is located on Node 5. The objective of this scenario is to evaluate the engine's ability to provide immersive user experiences. Notably, the direct 100Gbps link connecting the two nodes allows for an assessment of the performance metrics influenced by the engine's acceleration technology.

The outcome of these two scenarios will provide crucial insights into the performance and efficiency of the proposed architecture in managing operational data and its suitability for delivering immersive experiences. 
In both scenarios, to accurately measure the time taken for the transmission of the messages, the nodes are synchronized by using the Precision Time Protocol (PTP). This allows us to extract fine-grained metrics, and ensure that the results are accurate and reliable. 

\subsection{Results}

\begin{figure}[t]
  \centering
  \includegraphics[width=0.95\columnwidth]{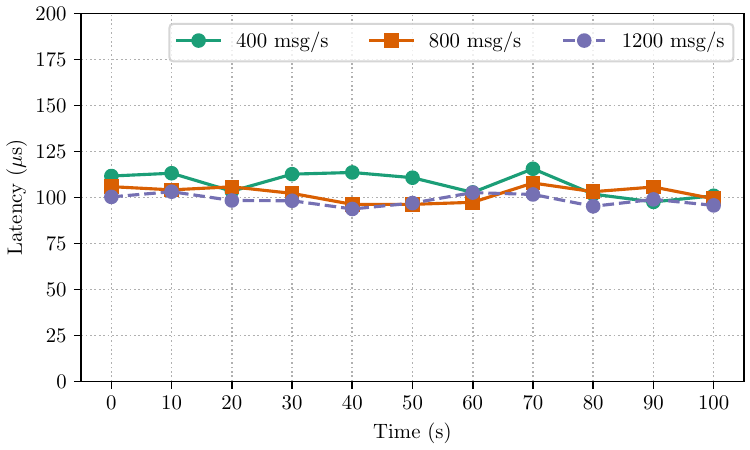}
  \caption{Machine-to-consumer communication latency under varying message load of the IT layer.}
  \label{fig:ot-latency}
\end{figure}

\begin{figure}[t]
  \centering
  \includegraphics[width=0.95\columnwidth]{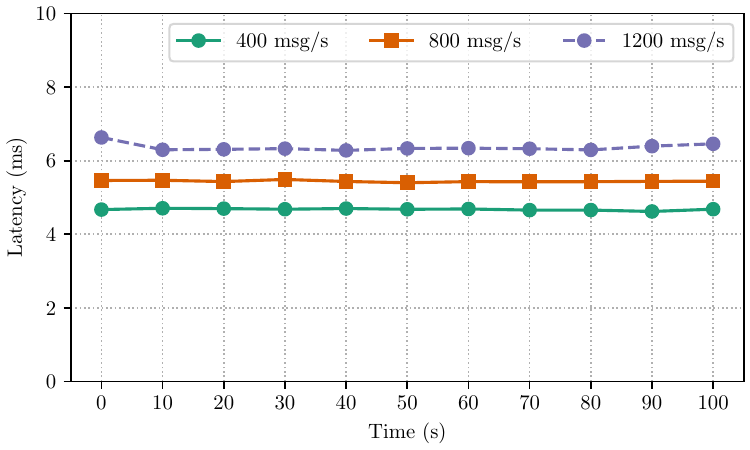}
  \caption{Machine-to-machine communication latency under varying message load of the OT layer.}
  \label{fig:it-latency}
\end{figure}

The proposed system's efficacy is assessed by measuring the message latency between the OT and the application layer, under varying traffic loads. The results, corresponding to the network observability application, are presented in Fig.~\ref{fig:ot-latency} and Fig.~\ref{fig:it-latency}.

Fig.~\ref{fig:ot-latency} illustrates the latency between two simulated machines, Node 1 and Node 2, in the OT layer, as the number of messages/second (real-time diagnostic information) is increased from \num{400} to \num{1200}. The latency is calculated as the time difference between receiving and sending messages at the application level. Our results show that the latency between the two machines remains stable and in the sub-millisecond range, which is the required latency for communication between machines or PLCs in the OT layer.


Fig.~\ref{fig:it-latency} displays the end-to-end latency between the OT layer and the Kafka consumer at the IT layer (Node 5) for the same message rate. The results show a latency that is an order of magnitude higher than the latency in the OT layer. This increase in latency is expected due to the number of software components the message must traverse, especially the latency introduced by the Kafka MOM topics, which have been configured to ensure a reliable and ordered delivery of messages conveyed from the OT. This configuration is particularly important for safety-critical data. The effects of this feature are more pronounced when increasing the number of messages/second, leading to an ever-increasing number of queued messages. To address this rate imbalance, the OT layer can be equipped with selective pre-processing capabilities, such as filtering and aggregation, reducing the burden at the OT/IT interface~\cite{icc2021}. 

It is worth noting that although the latency at the IT layer is higher than in the OT layer, it is still less than \SI{10}{\milli\second}, an acceptable threshold for many interactive applications, and sufficient to meet most use case requirements. Nevertheless, the fact that the latency grows significantly when the message rate increases highlights the need for further optimization and tuning, particularly in critical applications where low latency and data integrity are essential.

\begin{figure}[b]
  \centering
  \includegraphics[width=0.95\columnwidth]{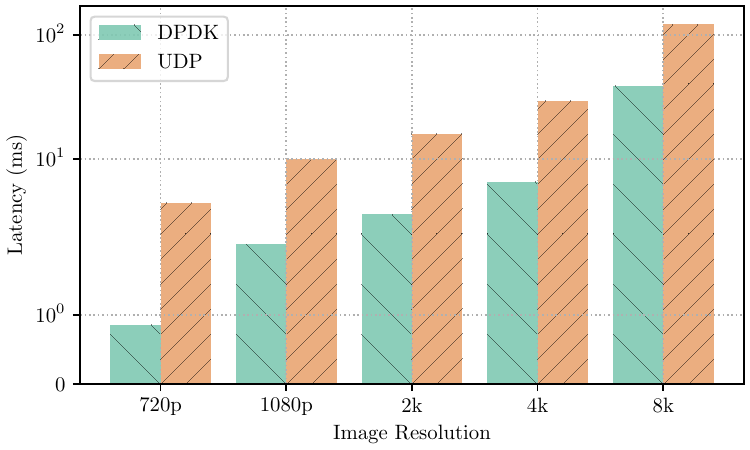}
  \caption{End-to-end latency for the streaming service as the resolution of the transmitted image increases (from HD to 8K).}
  \label{fig:streaming_latency}
\end{figure}

\begin{figure}[ht]
  \centering
  \includegraphics[width=0.95\columnwidth]{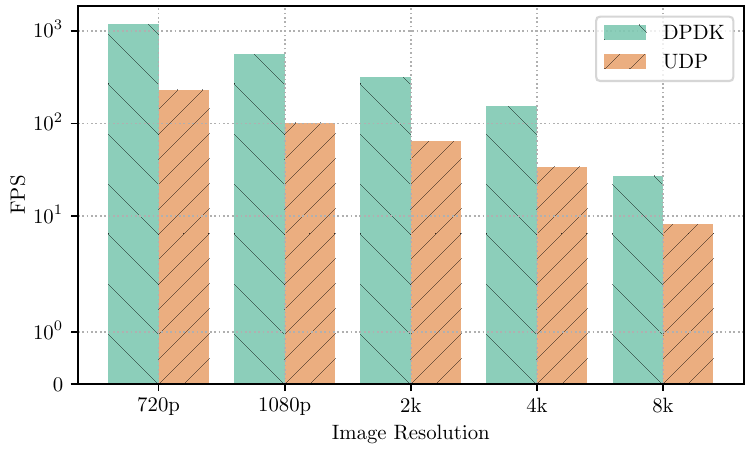}
  \caption{FPS for the streaming service as the resolution of the transmitted image increases (from HD to 8K).}
  \label{fig:streaming_fps}
\end{figure}

In the second experiment, we evaluate the performance of the streaming component (Metaverse engine) of the testbed by comparing two communication mechanisms for transmitting images of varying resolutions (from HD to 8K). One mechanism utilizes DPDK acceleration technology, while the other utilizes UDP traffic to simulate RTP-based communication~\cite{insanePoster}. We then analyze two performance metrics: (i) the number of frames per second (FPS) and (ii) the average end-to-end latency per frame transmission. The obtained results indicated an excellent performance of the streaming service in terms of both latency (Fig.~\ref{fig:streaming_latency}) and supported FPS (Fig.~\ref{fig:streaming_fps}), particularly in the case when relying on DPDK. The system can support frame rates above 100 FPS for images up to 4K resolution and exceed 1000 FPS for lower-quality images. Moreover, the observed end-to-end latency never exceeds \SI{10}{\milli\second} for images up to 4K resolution. These findings demonstrate that the system satisfies the requirements of next-generation interactive applications.

\section{Related Work}
\label{relatedwork}
To the best of our knowledge, the design and implementation of a comprehensive solution targeting Metaverse applications in the smart manufacturing domain are entirely novel in the existing literature. This section summarizes the prior research that influenced the authors' proposal, inspiring some architectural and technological choices.

The European Reference Architectural Model Industry 4.0 (RAMI 4.0)~\cite{rami} is a widely recognized standard that emphasizes the need for a close alignment between IT and OT. RAMI 4.0 provides a high-level reference architecture encompassing various Industry 4.0 scenarios. The reference communication layer of RAMI 4.0 leverages the OPC UA standard as the sole solution to ensure interoperability in the realm of OT~\cite{opc2}.

In \cite{survey_ind_ar}, the authors conducted a systematic literature review to assess the impact and effectiveness of using Augmented Reality (AR) in real industrial processes, investigating the applicability and usefulness of AR technology in the industry. The study found that AR is a growing trend and is being employed so to improve process flexibility, monitoring, and inspection capabilities, streamlining operations.

In~\cite{bosi2019cloud}, the authors propose a multi-layer architecture for monitoring industrial equipment in customer plants. The architecture uses two Apache Kafka installations, one in OT and one in IT, to gather near-real-time data. Dedicated software components called \emph{hmi-forwarders} interface with Modbus-TCP machinery, exporting data to the OT Kafka instance, which then forwards it to the appropriate Kafka topic in the IT layer.

The CODEG framework for cloud-oriented gaming, proposed by DeGiovanni et al.~\cite{de2022revamping}, addresses the limitations of conventional monolithic game engines, advocating for a distributed framework that can exploit the full potential of the heterogeneous resources in the cloud continuum. The implementation of the CODEG framework involves the integration of multiple game engine modules into the available network infrastructure, ranging from core to edge resources.

\section{Conclusion and Future Work}
\label{conclusion}
In this article, we presented a conceptual architecture and discussed the accompanying software prototype and functional components. The proposal aims at enabling an immersive smart manufacturing environment, embodying a human-in-the-loop approach. A preliminary evaluation was presented, validating the framework components as a whole by relying on two proof-of-concept immersive application scenarios. Our ongoing efforts are directed toward expanding the Metaverse engine and the applications described in this study to enable the creation of shared virtual environments where multiple users can engage in synchronous sessions.



\bibliographystyle{IEEEtran}
\bibliography{IEEEabrv,main}

\end{document}